\documentstyle[twocolumn,aps,epsf]{revtex} %galley
\begin{document}
%-----------------------
\title{Measuring Polarized Gluon and Quark Distributions
with Meson Photoproduction}
%-----------------------
\author{
Andrei Afanasev{\footnotemark}
}
%----------------------------
\address{
Thomas Jefferson National Accelerator Facility,
12000 Jefferson Avenue, Newport News, VA 23606\\
and Department of Physics, Hampton University, Hampton,
VA 23668}
%------------------------------
\author{
Carl E. Carlson and Christian Wahlquist
}
\address{
Physics Department, College of William and Mary,
Williamsburg, VA 23187
}
%-------------------
\date{revised \today}
%------------------
\maketitle
%-------------------
\begin{abstract}
%\parshape=1 0.75in 5.5in \indent
%{\small 
We calculate polarization asymmetries in
photoproduction of high transverse momentum mesons, focusing on
charged pions, considering the direct, fragmentation, and resolved
photon processes.  The results at very high meson momentum
measure the polarized quark distributions and are sensitive
to differences among the existing models. The results at
moderate meson momentum are sensitive to the polarized gluon
distribution and can provide a good way to measure it. 
Suitable data may come as a by-product of deep inelastic
experiments to measure $g_1$ or from dedicated
experiments.       %\vglue -20pt}
\end{abstract}

%\widetext \vglue -6.75cm  
%{\sl DRAFT} 
%\hfill JLAB-THY-97-27, WM-97-104, hep-ph/9706522
%\quad {\sl DRAFT}
%\vglue 6cm \narrowtext
%-------------------------

%-----------------

\footnotetext{\vglue -25pt $^*$On leave from Kharkov
Institute  of Physics and Technology, Kharkov, Ukraine.}

%-------------------------------------------

\section{Motivation}

In this article, we will discuss photoproduction of high
transverse momentum pions from polarized initial states. 
There are three motivations for doing so.  One is the
opportunity to learn the polarization distribution of quarks
and gluons in nucleons.  We will show that pion
photoproduction with polarized initial states has, over a
wide kinematic region of moderate transverse momentum pions, a
large sensitivity to the polarized gluon distribution
functions of the target.  Within this wide kinematic region
there are broad circumstances where the known polarized quark
distributions all give similar pion photoproduction
contributions, so that differences among the results are due
mainly to the polarized gluon distributions.  Hence data
where rather ordinary mesons are produced can select among
the various models for this quantity.

Another motivation, which we have written something about
earlier~\cite{acw1}, comes from the kinematic region of very
high transverse momentum pions where the gluon contributions
are small but the differences among the various models for the
polarized quark distributions are significant.  Hence
examining different kinematic regions of pion photoproduction
yields information about both polarized quark and
polarized gluon distributions.

A third motivation, also dependent upon the highest
transverse momentum pions, is the possibility of learning
something about the pion distribution amplitude.  In this
region,  ratios of cross sections determine the target's
quark distributions, but the magnitude of the cross section
depends upon the same integral involving the pion distribution
amplitude that enters the pion electromagnetic form factor of
the $\pi^0 \gamma \gamma^*$ vertex.  Hence if one looks at
the unpolarized case, where the target distributions are
fairly well known, one has another measure of this
integral.  (Pion photoproduction in the unpolarized case has
been well and successfully studied theoretically and
experimentally~\cite{aurenche}, but not at the highest
transverse momenta where the pions will be dominantly
produced in a short distance process rather than via
fragmentation~\cite{acw1,cw93,bgh}.) 

Presently, information on polarized quark distributions
comes from deep inelastic electron or muon scattering
with polarized beams and targets~\cite{g1}.  Single arm
measurements of $g_1$ give information about a
charge-squared weighted combination of polarized quark
distributions. Obtaining polarized distributions of individual
flavors is not possible from this data alone, but requires
extra theoretical input in the analysis.  Coincidence
measurements of
$\vec \ell\ \vec p (\vec d) \rightarrow \ell\, \pi^\pm X$
give more information and have been reported~\cite{adeva}.
This data, for a proton or deuteron target, gives different
linear combinations of up and down quark polarized
distributions, allowing a flavor decomposition without
further theoretical input~\cite{flavor}.

\begin{figure}

\vglue -0.1in
\epsfysize 2.7 in \epsfbox{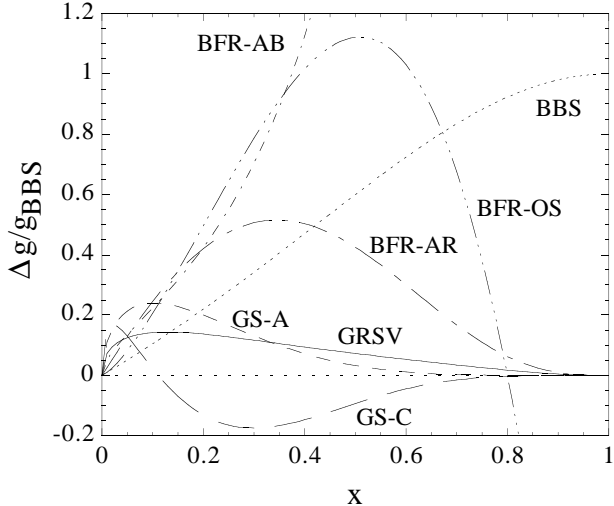}
\vglue 0.05 in

\caption{A number of polarized gluon distributions, all
normalized to the unpolarized BBS gluon distribution.
The sources of these distributions are given later in the text.}

\label{polglue7}
\end{figure}

Polarized gluon distributions are not well determined
at present.  Something is learned
from~\cite{bbs,gs96,grsv96,bfr96} the measurements of
$g_1$, but gluons contribute to $g_1$ only in higher order or
through their effects upon the evolution of the polarized
quark distributions.  The analyses of $g_1$ can be abetted by
perturbative QCD considerations at high $x$~\cite{bbs}. 
Overall, however, the present constraints upon the polarized
gluon distributions are not great and there is a
large variance among $\Delta g(x,\mu^2)$ models,  as may be seen
in Fig.~\ref{polglue7}.

The process we discuss,
$\vec \gamma\, \vec p \rightarrow \pi \, X$ (where the
photon is real and targets other than protons are possible),
gives a complementary way to find the polarized quark
distributions and is sensitive to the gluon distributions in
leading order.  The perturbative QCD that we use in the analysis
is justified on the basis of high meson transverse momentum,
rather than by high virtuality of an exchanged photon,  and the
experiment is a single arm experiment rather than a coincidence
one.  Good data can in fact come as a by-product of a $g_1$
experiment since the detectors that measure the final electron or
muon can also pick up charged hadrons;  recall that if the final
lepton is not measured, the form of the cross section ensures
that the virtuality of the exchanged photon will in general be
rather low.

%%%%%%%%%%%%%%%%%%%
%%%%%%%%%%%%%%%%%%%

There are several processes that produce pions.  At the
highest transverse momenta, mesons are produced by short
range processes illustrated in Fig.~\ref{direct}. We call
these direct processes because the photon interacts
directly with the target partons and also the pion is
produced immediately.  (The word ``direct'' was used with a
similar meaning in a $\pi N \rightarrow \gamma^* X$ context by
Brodsky and Berger long ago~\cite{bb81}.)   Direct processes
are amenable to perturbative QCD calculation~\cite{cw93,bgh} and
produce mesons that are kinematically isolated in the direction
they emerge.  These processes possess several nice features.  One
is that if the pion three-momentum is measured, no integrals are
needed to calculate the differential cross section. In
particular, the momentum fraction $x$ of the struck quark is
fixed by measurable quantities.  Formulas for
$x$ were given in~\cite{acw1} and will be repeated below,
this time including mass corrections.  The
situation is reminiscent of deep inelastic lepton
scattering, where experimentally measurable quantities,
$Q^2$ and $\nu$, determine the quark momentum fraction by
$x = Q^2/2m_N \nu$.  Another nice feature is that the
asymmetry for the meson production subprocess is easy to
calculate and is large.   Finally, mesons of a given flavor
come mainly from quarks of a given flavor.  Hence in the
region where the direct process dominates, we can choose which
flavor quark we study the polarization distribution of.  Note
that the high transverse momentum region involves only the
high $x$ quarks and that the various models for the
quark distributions separate from each other at high $x$.

At moderate pion transverse momentum, the main process is
one we call the fragmentation process.  The photon does
interact directly with the partons of the target, but the
meson is produced by fragmentation of one of the final
state quarks or gluons.  This time, an integration is needed
to calculate the cross section, but any given model makes a
definite prediction that can be compared to data.  Unlike the
case for $g_1$, interactions involving the gluons in the
target contribute to the cross section in lowest order.  One
of the important subprocesses is photon-gluon fusion, 
$\gamma + g \rightarrow q + \bar q$. The
polarization asymmetry of this process is very large. 
Indeed, neglecting masses, it is
$-100$\%:  the process only goes if the photon and gluon have
opposite helicity.  Hence this process is even more
significant for the polarization asymmetry than it is
overall.  There are situations where the results excluding
the gluon polarization are close to the same for all the
modern parton distribution function models or
parameterizations.  Then the differences among the results
from different models are due to the polarized gluon
distributions, and the differences over the spectrum of
available models are large.  Hence, the data will
adjudicate among the  different suggest polarized gluons
distributions.  

There is also the resolved photon process, where the photon
turns into hadronic material before interacting with the
target. We will discuss it in some detail below.  However,
for the kinematic situations we highlight, the resolved
photon contributions are below both the fragmentation and
direct contributions.

Calculational details are outlined in the following
section.  Some results and tests of the calculations are
outlined in the next following section.  Then, in 
section~\ref{polarized}, we show results involving polarized
initial states, in particular showing how sensitive the
results are the the different models for the polarized
parton distributions and how well they could be extracted
from the data.  Some conclusions will be given in
section~\ref{discussion}.

%-------------------------------------------

\section{Calculations}  \label{calculations}

There are three categories of processes that contribute to pion
photoproduction and we call them fragmentation
processes, direct processes, and resolved photon processes.  

Fragmentation processes have quarks and gluons produced in short
range reactions followed by fragmentation at long distances of
either a quark or a gluon to produce the observed pion.  The short
distance part of the process is perturbatively calculated and the
long distance part is parameterized as a fragmentation function
for partons into pions.  Direct processes, in our nomenclature,
occur when the pion is produced in a short range reaction via a
radiated gluon giving a quark-antiquark pair, one of which joins
the initial quark to produce the pion.  This process is
perturbatively calculable, given the distribution of initial
quarks, and produces isolated pions rather than pions as part of
a jet.  The direct process can dominate the fragmentation
process for very high transverse momentum pions.  Resolved photon
processes are photons fluctuating into hadrons, most simply a
quark-antiquark pair, which then interact with the partons of the
target.  The resolved  photon processes can be important for high
initial energy, especially for pions produced backward in the
center of mass.

\begin{figure}

%\vglue 0.in
\hskip 0.5 in \epsfysize 1.2 in \epsfbox{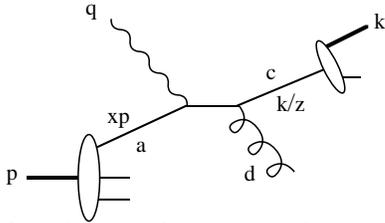}

\caption{One diagram for photoproducing $\pi$ mesons via
fragmentation.}

\label{fragmentation}
\end{figure}

Fragmentation processes, of which one example is shown in
Fig.~\ref{fragmentation}, are important over a wide range of
kinematics for the present paper, and we start by recording the
relevant formulas~\cite{peralta}.  In general, if the photon
interacts directly with a constituent of target $N$ but the pion
is produced as part of a jet,
%%%%%%%%%%%%%%%%%%%
%%%%%%%%%%%%%%%%%%%

%
\begin{equation}
\sigma = \sum_{a,c,d} \int G_{a/N}(x) 
   {d\hat\sigma \over d \hat t}(\gamma + a \rightarrow c + d)
   D_{\pi/c}(z) \, dx \, d \hat t \, dz.
\end{equation}
Here, $x$ is the (light-cone) momentum fraction of the target
carried by the struck parton $a$, $z$ is the fraction of the
parton $c$'s momentum that goes into the pion, and $\hat s$,
$\hat t$, and $\hat u$ are the Mandelstam variables for the
subprocess $\gamma + a \rightarrow c + d$.  The scale dependence
of the parton distribution functions $G$ and the fragmentation
functions $D$ will often be tacit, as it is above.  As a
differential cross section for the pion, one gets
\begin{eqnarray}
E_\pi {d\sigma \over d^3k} &=& {s-m_N^2 \over -\pi t} 
  \sum_{a,c,d} \int_{z_{min}}^1  {dz \over z} x^2  \times
\nonumber \\[1.5ex]
 &&  G_{a/N}(x) 
   {d\hat\sigma \over d \hat t}(\gamma + a \rightarrow c + d)
   D_{\pi/c}(z)   ,
\end{eqnarray}
The Mandelstam variables for the overall process $s$, $t$, and
$u$ are defined for the inclusive process by
\begin{eqnarray}
s = (p+q)^2 \nonumber \\
t = (q-k)^2 \nonumber \\
u = (p-k)^2
\end{eqnarray}
where $q$, $p$, and $k$ are the momenta of the incoming photon,
the target, and the outgoing pion, respectively.  The lower
integration limit is
\begin{equation}
z_{min} = - { t + u - m_N^2 \over s - m_N^2 } ,
\end{equation}
and 
\begin{equation}
x = { - t \over z( s - m_N^2 ) + (u - m_N^2) }  .
\end{equation}

When the target and projectile are polarized, we define
\begin{equation}
\Delta \sigma ={1\over 2}
   \left( \sigma_{R+} - \sigma_{R-} \right) ,
\end{equation}
where $R$ and $L$ represent photon helicities and $\pm$
represent target helicities, and similarly for $\hat \sigma$. 
Also, the polarized parton distributions are defined by
\begin{equation}
\Delta G_{a/N}(x) = \Delta G_{a/N}(x,\mu^2) =
  G_{a+/N+}(x) - G_{a-/N+}(x).
\end{equation}
For quarks and gluons and proton targets we will often use the
notation $q(x) \equiv G_{q/p}(x)$ and $g(x) \equiv G_{g/p}(x)$ and
their polarized equivalents.  The cross section is now given by
\begin{eqnarray}
E_\pi {d\Delta\sigma \over d^3k} &=& {s-m_N^2 \over -\pi t} 
   \sum_{a,c,d} \int_{z_{min}}^1  {dz \over z} x^2  \times
\nonumber \\[1.5ex]
 &&  \Delta G_{a/A}(x) 
{d\Delta\hat\sigma \over d \hat t}(\gamma + a \rightarrow c + d)
   D_{\pi/c}(z)   .
\end{eqnarray}

The relevant subprocess cross sections are
\begin{eqnarray}
{d\hat\sigma \over d\hat t}(\gamma+q \rightarrow g+q) &=& 
  {8\pi e_q^2 \alpha\alpha_s \over 3 \hat s^2}
\left( {-\hat s \over \hat u} + {\hat u \over -\hat s} \right),
                         \nonumber \\
{d\hat\sigma \over d\hat t}(\gamma+g \rightarrow q + \bar q)
                       &=& 
  {\pi e_q^2 \alpha\alpha_s \over  \hat s^2}
 \left( {\hat u \over \hat t} + {\hat t \over \hat u} \right),
                         \nonumber \\
{d\Delta\hat\sigma \over d\hat t}(\gamma+q \rightarrow g+q) &=& 
  {8\pi e_q^2 \alpha\alpha_s \over 3 \hat s^2}
\left( {\hat s \over -\hat u} - {\hat u \over -\hat s} \right),
                         \nonumber \\
{d\Delta\hat\sigma \over d\hat t}
     (\gamma+g \rightarrow q + \bar q)
                       &=& -
  {\pi e_q^2 \alpha\alpha_s \over  \hat s^2}
 \left( {\hat u \over \hat t} + {\hat t \over \hat u} \right).
\end{eqnarray}
The cross section for $\gamma+q \rightarrow g+q$ is written
for $\hat t$ being the momentum transfer between the photon
and the gluon. The asymmetry for the quark target is
positive and the asymmetry for the gluon target is $-100$\%.

%%%%%%%%%%%%%%%%%%%
\begin{figure}[t]

\vglue -0.1in
\hskip 0.9 in \epsfysize 1.1 in \epsfbox{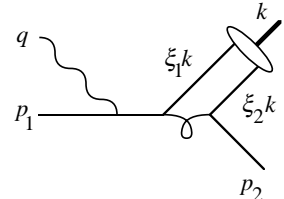}

\caption{One of four lowest order perturbative diagrams for
direct photoproduction of mesons from a quark.  The four
diagrams correspond to the four places a photon may
be attached to a quark line.}

\label{direct}
\end{figure}
%%%%%%%%%%%%%%%%%%%

For the direct process, the subprocess is shown in
Fig.~\ref{direct}. When the incoming photon is
circularly polarized and target quark is longitudinally
polarized, one gets to lowest order
\begin{eqnarray}
{d\hat\sigma(\gamma q \rightarrow \pi q') \over dt} &=&
  {128 g_F^2 \pi^2 \alpha \alpha_s^2 \over 27 (-t) \hat s^2}
    I_\pi^2
  \left({e_q \over \hat s} + {e_{q'} \over \hat u} \right)^2
     \nonumber \\[1ex] &\times&
\left[ \hat s^2 + \hat u^2
                 + \lambda h
       \left( \hat s^2 - \hat u^2
       \right)
    \right],
\label{directeqn}
\end{eqnarray}

\noindent where $\lambda$ is the helicity of the photon,
$h$ is twice the helicity of the target quark, and $g_F$ is a
flavor factor from the overlap of the $q \bar q'$ with
the flavor wave function of the meson.  It is unity for
most mesons if the quark flavors are otherwise suitable;
for example
\begin{equation}
g_F = \left\{
\begin{array}{cc}
1/\sqrt{2}  &  {\rm for\ } \pi^0 \\
1          &  {\rm for\ } \pi^+
\end{array}   \right. .
\end{equation}

\noindent The integral $I_\pi$ is given in terms of the
distribution amplitude of the meson
\begin{equation}
I_\pi = \int {d\xi_1 \over \xi_1} \phi_\pi(\xi,\mu^2).
\end{equation}
It is the same integral which appears in
the perturbative calculation of the $\pi^\pm$
electromagnetic form factor or of the $\pi^0 \gamma \gamma$
form factor.  For the asymptotic distribution amplitude,
$\phi_\pi(\xi) = 6\xi_1 \xi_2 \cdot f_\pi/2\sqrt{3}$, one gets
$I_\pi = \sqrt{3} f_\pi/2$, where $f_\pi \approx 93 MeV$.
Finally, in this case
\begin{eqnarray}
\hat s = (p_1 + q)^2, \nonumber \\
\hat u = (p_1 - k)^2,
\end{eqnarray}
and $\hat t$ is the same as $t$.

The direct process is higher twist, nominally suppressed by a
factor of scale $f_\pi^2/s$.  However, for very high
transverse momentum pions it is the dominant
production process.  When it is the dominant process, one can
take advantage of the nice feature that the momentum fraction
$x$ of the struck quark is completely determined by
experimentally measurable quantities. With $p_1 \simeq x p$
and estimating mass corrections with a proportional mass
approximation, one has
\begin{equation}
\hat s + \hat t + \hat u = 2x^2 m_N^2.
\end{equation}
Further,
\begin{equation}
\hat s = x s - x(1-x) m_N^2, \quad
\hat t = t,   \quad
\hat u = x u - x(1-x) m_N^2.
\end{equation}
Hence,
\begin{equation}
x ={-t \over s + u - 2 m_N^2}.        
\label{x}
\end{equation}

Thus to the overall process, the direct subprocess makes a
contribution that requires no integration to evaluate.  For
the polarized case
\begin{eqnarray}
&& E_\pi {d\Delta\sigma \over d^3k} = 
  {(s - m_N^2) x^2 \over - \pi t}
  {d\sigma(\gamma p \rightarrow \pi + X) \over dx\, dt}
\nonumber  \\
 && \  = 
    {(s - m_N^2) x^2 \over - \pi t}  
    \sum_q \Delta G_{q/p}(x,\mu^2)
    {d\Delta\hat\sigma(\gamma q \rightarrow \pi q') \over dt},
\end{eqnarray}
where the helicity summations are tacit.  The unpolarized case
is the same with the $\Delta$'s.

\begin{figure}

%\vglue 0.in
\hskip 0.5 in \epsfysize 1.2 in \epsfbox{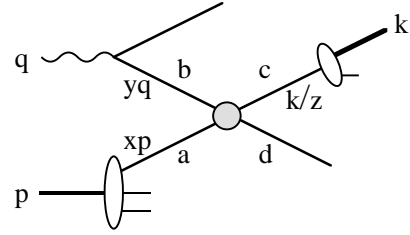}

\caption{A resolved photon process.}

\label{resolved}
\end{figure}

We now come to the resolved photon contributions.  One such
contribution is illustrated in Fig.~\ref{resolved},  where
the photon turns into hadrons such as a quark-antiquark pair
before interacting with the target.

%%%%%%%%%%%%%%%%%%%
%%%%%%%%%%%%%%%%%%%

The cross section is
\begin{eqnarray}
E_\pi {d \Delta \sigma \over d^3k} &=& \sum_{abcd} 
   \int_{y_{min}}^1 dy  \int_{z_{min}}^1  dz \,
  {x^2 (s-m_N^2) \over -\pi z t} 
                                   \nonumber \\[1.4ex] 
 &\times& \Delta G_{b/\gamma}(y)  \,  \Delta G_{a/N}(x)
  {d \Delta \sigma \over d\hat t} D_{\pi/c}(z),
\end{eqnarray}
where again the scale dependence is tacit and the
unpolarized case is got by removing the $\Delta$'s.  Also,
\begin{eqnarray}
x &=& {-yt \over yz(s-m_N^2) + (u-m_N^2)} ,
        \nonumber \\
y_{min} &=& { -(u-m_N^2) \over (s-m_N^2) + t } ,
        \nonumber \\
z_{min} &=& { -(u-m_N^2) - yt \over y(s-m_N^2) } .
\end{eqnarray}

For estimating the size of the resolved photon contributions,
we  began with the lowest order non-trivial result for the
photon splitting function~\cite{peralta,hp81}, 
\begin{equation}   \label{perturbative}
G_{b/\gamma}(y) = 3e_b^2 {\alpha \over 2\pi}
   \left[y^2 + (1-y)^2 \right] 
    \ln \left( Q^2 \over Q_0^2 \right) 
\end{equation}
and
\begin{equation}
\Delta G_{b/\gamma}(y) = 3e_b^2 {\alpha \over 2\pi}
  \left[y^2 - (1-y)^2 \right] 
    \ln \left( Q^2 \over Q_0^2 \right) ,
\end{equation}
 with $Q_0 = 0.3$ GeV.  In addition, we
considered the more complete parametrizations~\cite{gamma} that
use vector meson dominance to initialize the non-perturbative
parton densities.

The subprocess cross sections, both polarized
and unpolarized are available in~\cite{babcock79}.  For a
quark or antiquark scattering off a gluon in the target, 
\begin{eqnarray}
{d(\Delta)\sigma \over d\hat t}
(qg \rightarrow qg) =
  {\pi \alpha_s^2 \over \hat s^2}  
   \left( \hat s^2 \pm \hat u^2 \right)
   \left( 
{1\over \hat t^2} - {4\over 9 \hat s  \hat u}
                                        \right)  ,
\end{eqnarray}
where the upper and lower signs correspond to the unpolarized
and polarized cases, respectively, and $\hat t$ is the
momentum transfer from incoming to outgoing quark.  For
quark-quark (or antiquark-antiquark) scattering,
\begin{eqnarray}
{d(\Delta)\sigma \over d\hat t}
(q_\alpha q_\beta \rightarrow q_\alpha q_\beta) &=&
  {\pi \alpha_s^2 \over \hat s^2}  {4\over 9}
   \biggl\{ {\hat s^2 \pm \hat u^2 \over \hat t^2} 
                           \nonumber \\
&+&
   \delta_{\alpha\beta} 
   \left[  {\hat s^2 \pm \hat t^2 \over \hat u^2}
         -{2 \hat s^2 \over 3 \hat t \hat u}
                                               \right]
                                               \biggr\}  ,
\end{eqnarray}
where the coding of the $\pm$ is the same as above, the
subscripts on $\delta_{\alpha\beta}$ refer to flavors of
incoming quarks and $\hat t$ is the momentum transfer between
incoming and outgoing quarks of the same flavor. (Note, of
course, the  $\hat t \leftrightarrow \hat u$ symmetry for the
all same flavor case.)  The last case is scattering of
quark by antiquark,
\begin{eqnarray}
{d(\Delta)\sigma \over d\hat t}
(q_\alpha \bar q_\beta &\rightarrow& q_\delta \bar q_\gamma) 
=
  {\pi \alpha_s^2 \over \hat s^2}  {4\over 9}
   \biggl\{ 
     \delta_{\alpha\delta}  \delta_{\beta\gamma}
     {\hat s^2 \pm \hat u^2 \over \hat t^2} 
                           \nonumber \\
&\pm&
   \delta_{\alpha\beta} \delta_{\gamma\delta}
   \left[  {\hat t^2 + \hat u^2 \over \hat s^2}
         - \delta_{\alpha\delta}
           {2 \hat u^2 \over 3 \hat s \hat t}
                                               \right]
                                               \biggr\}  ,
\end{eqnarray}
where still again the upper and lower signs correlate with
the unpolarized and polarized cases, $\hat t$ is the
momentum transfer from incoming to outgoing quark, and
$\alpha, \beta, \gamma$, and $\delta$ are flavor indices.

Good data can come from electroproduction experiments
where only the outgoing pion is observed.  Because of the
$q^{-4}$ in the cross section, the photons are nearly all
close to real, and the equivalent photon approximation
gives the general connection between the electroproduction
and photoproduction cross sections~\cite{bkt71},
\begin{eqnarray}
d\sigma(eN \rightarrow \pi X) =
  \int_{E_{min}}^{E_e} dE_\gamma \,
      N(E_\gamma) d\sigma(\gamma N \rightarrow \pi X).
\end{eqnarray}
where $E_\gamma$ is the energy of the photon and
\begin{eqnarray}
      &N(E_\gamma)& =  {\alpha\over \pi E_\gamma}
\biggl[   {E_e^2 + {E'_e}^2 \over E_e^2} 
          \left( \ln{E_e \over m_e}-{1\over 2} \right)
+ {E_\gamma^2\over 2 E_e^2} \times
\nonumber \\[1.7ex]  &\times&
          \left(  \ln {2 E'_e\over E_\gamma} +1  \right)
+   {(E_e + E'_e)^2 \over 2 E_e^2} 
            \ln {2{E'}_e\over E_e + E'_e}      \biggr]  ,
\end{eqnarray}
where $E'_e = E_e - E_\gamma$. The lower limit on the photon
energy integral is
\begin{equation}
E_{min} = {k \over 
  1 - 2(k / m_N) \sin^2 (\theta_{lab} / 2)}  .
\end{equation}

When the electron is polarized, the polarization transfers
nicely to the photon provided the photon takes most of the
electron's energy.  Polarization details can be found
in~\cite{hm}; most usefully, if
$P_\gamma$ and $P_e$ are the circular and longitudinal
polarizations of the photon and electron, respectively, then
\begin{equation}
{P_\gamma \over P_e} = {y'(4-y') \over 4-4y'+3y'^2}.
\end{equation}

\noindent where $y' = E_\gamma/E_e$.

We close this section with a few comments on our procedures. 

We used $\alpha_s(\mu^2)$ with the renormalization scale set
to the pion transverse momentum.  Not all the pieces needed
for are calculation are known beyond leading order and we have
worked to lowest order throughout. We took $\Lambda_{QCD} =
175$ MeV for four flavors.  (This corresponds, to about
$\pm\, 6$\% in $\alpha_s$ for $\mu = 1-5$ GeV, to a four flavor
$\Lambda_{QCD}$ of 295 MeV in the next to leading order
formula, which matches to a five flavor $\Lambda_{QCD}$ of 209
MeV, which is the central value quoted by the Particle Data
Group~\cite{pdg}.  Uncertainties are roughly $\pm 40$ MeV on
the 209 and 175 MeV numbers, and $\pm 50$ MeV on the 295 MeV.)

The fragmentation functions we use may be found in the
Appendix of~\cite{cw93}.    Briefly, the
fragmentation of quarks into pions contains a part $D_p$ if
the primary quark is a valence quark in the pion, and also a
secondary part
$D_s$ for any quark-pion combination.  Three examples are
\begin{eqnarray}
D_{\pi^+/u} &=& D_p + D_s,\quad 
D_{\pi^0/u} ={1\over 2} D_p + D_s, \nonumber \\
D_{\pi^-/u} &=&  D_s  .
\end{eqnarray}
At the benchmark scale (which we took to be 29 GeV),
\begin{eqnarray}
D_p = {5\over 6} (1-z)^2 \quad {\rm and} \quad
D_s = {5\over 6} {(1-z)^4 \over z}.
\end{eqnarray}
These forms lead to good fits to the 
$e^+e^- \rightarrow \pi + X$ data, and so we stick with them.
At the time of reference~\cite{cw93}, the known fragmentation
functions were more than a decade old and no longer fit
up-to-date data.  Since then a number of other modern
fragmentation functions have appeared~\cite{bkk95}, which
also match data.  The benchmark gluon fragmentation function
is 
\begin{equation}
D_{\pi/g} = {2\over 3} {(1-z)^3 \over z}.
\end{equation}

The mass corrections we estimated using a proportional mass
approximation.  We gave the parton that came from the target
a mass $x m_N$, and gave the final parton that did not go
into the pion the same mass.  The parton that did go into the
pion we treated as massless (like the pion), and did the same
for the parton that came from the photon in the resolved
photon process.  A fully defensible treatment of mass
corrections would require a solution to QCD.  The proportional
mass approximation just described has the virtues of being
simple and of giving the same kinematic limits from
thresholds and energy conservation for the subprocess as for
the overall process.  Hence, it is an improvement over
putting in no mass corrections, though we may treat it
largely as a way to receive a warning to be careful when the
mass corrections are big.  For the situations we study, the
mass corrections are not large except when the cross sections
are very small.

%-------------------------------------------

\section{Results without Polarization}  \label{unpol}

%-------------------------------------------

\begin{figure}
\hglue 0.1 in \epsfysize 2.24 in \epsfbox{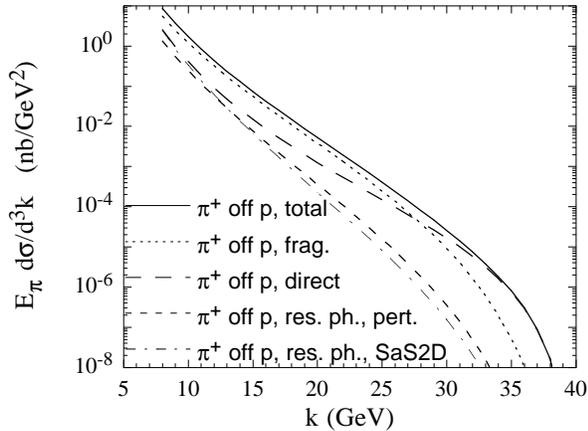}

\vglue 0.05 in

\caption{Comparing fragmentation, direct, and resolved
photon processes for 
$e + p \rightarrow  \pi^+ + X$ with
$E_e = 50$ GeV and $\theta_{lab} =
5.5^\circ$.  These all use the GRSV parton
distributions.  The relative
size of the contributions should not depend much on which
parton distributions we use.  For the resolved photon
contribution, the result of both the perturbative splitting
function for the photon, eqn.~(\protect\ref{perturbative}), and
the more sophisticated SaS 2D are shown.  Both are small for
these kinematics.}

\label{compare50}
\end{figure}

Our present main interest is on results obtainable
for polarized beams and targets.  However, both for checking
the model and for intrinsic interest we will present some
results with no polarization involved.  First in
Fig.~\ref{compare50} we show the relative size of the
fragmentation, direct, and resolved photon contributions for
some kinematics of interest, namely 50 GeV electrons with
the only observed final state particle being a $\pi^+$
emerging at 5.5$^\circ$ in the lab.  The resolved photon
curve used the simple perturbative parton densities in the photon
given by eqn.~(\ref{perturbative}).  The resolved contributions
are still small for our kinematics if one of the more realistic
distributions be used.   As an example, Fig.~\ref{compare50}
also shows the resolved photon contribution using SaS 2D from the
last of ref.~\cite{gamma}.  It increases the resolved photon
result by about 80\% for the smallest momentum shown and then
gives a smaller result for momenta above 13 GeV. In the 
kinematics of Fig.5, average fractions of the photon momentum carried by its partons
are $<y>$=0.69 for the lowest 
pion momentum in Fig.5, 0.8 for momentum 15 GeV and 0.9 for 
momentum 24 GeV. Because of these large values of $<y>$, we neglected a possible
gluon fusion process ($gg\to q\bar q$).

Commentary on the quark and gluon distribution models we
use is put in the next section, so that we can bundle the
remarks on the polarzed and unpolarized distributions into one
location.

%%%%%%%%%%%%%%%%%%%
%%%%%%%%%%%%%%%%%%%

One sees that the cross section falls quickly with
increasing pion momentum.   The fragmentation process is the main one at lower pion
momenta, and the direct process takes over above about 26 GeV
for this particular angle and incoming energy.

In addition, one sees that the resolved photon process is
not particularly important here.  At higher energies it
increases in relative importance~\cite{xxx} and at HERA energies
($\sqrt{s} \approx 300$ GeV), the resolved photon process
dominates except for very forward angles.  The reason for its
fast increase involves the lower average $y$ possible at higher
energies, as well as the reduced kinematic constraint upon a
three step (three integrals in the calculation) process as the
energy increases.

\begin{figure}

\hglue 0.1in \epsfysize 2.25 in \epsfbox{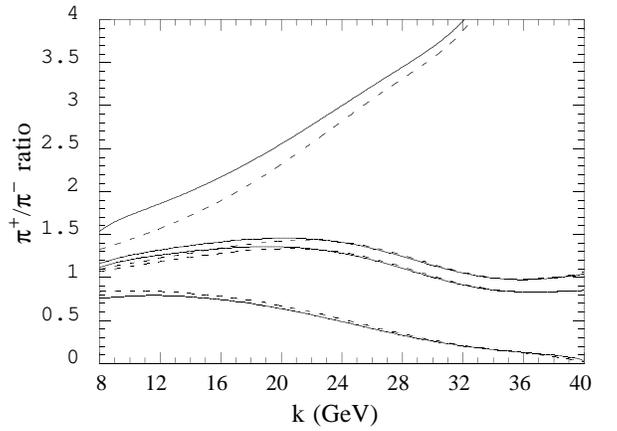}

\vglue 0.05in

\caption{The $\pi^+/\pi^-$ ratio for $E_e = 50$ GeV and
$\theta_{lab} = 5.5^\circ$ for the pions.  The dashed curve
is for GRV and the solid curve is for CTEQ.  Both have
$u(x)/d(x) \sim 1/(1-x)$ for large $x$.  The four pairs of
curves are for, top to bottom, protons, 
a target which is 5/9 protons and 4/9 neutrons,
an isoscalar target, and neutrons.}

\label{pionrat50}
\end{figure}

\begin{figure}

\hglue 0.1 in \epsfysize 2.3 in \epsfbox{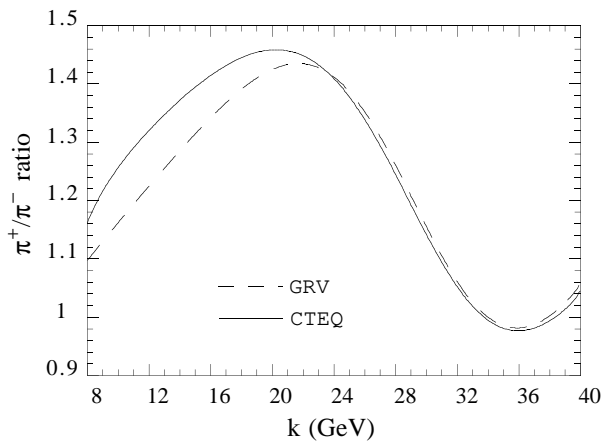}

\vglue 0.05in

\caption{Detail of $\pi^+/\pi^-$ ratio for $E_e = 50$ GeV and
$\theta_{lab} = 5.5^\circ$ for the pions, for a target which is
$5/9$ protons and 4/9 neutrons.}

\label{pionratio50}
\end{figure}

The $\pi^+/\pi^-$ ratio off proton and neutron targets is shown
in Fig.~\ref{pionrat50}, again for 50 GeV incoming
electrons.  Most of the models for the parton distributions
lead to the similar results except at the highest $k$, which will
be discussed below.  Also included in the figure are two sets of
predictions for the $\pi^+/\pi^-$ ratio of isoscalar or nearly
isoscalar targets.
%%%%%%%%%%%%%%%%%%%
%%%%%%%%%%%%%%%%%%%
We detail in Fig.~\ref{pionratio50} the $\pi^+/\pi^-$ ratio
for 50 GeV incident electrons on a target that is $5/9$
protons and 4/9 neutrons.  (These are relevant numbers for
one actual ammonia target, where the nitrogen is $^{15}$N and
the hydrogens have plain proton nuclei.) 

The predictions for the $\pi^+/\pi^-$ ratio are different for
purely direct and purely fragmentation processes.  For
an isoscalar target, the ratio is not sensitive to the quark
distributions, and the observed behavior of the $\pi^+/\pi^-$
ratio could be a clear signal for the direct process taking over
from fragmentation with increasing pion momentum. For a proton
or neutron target, there is much sensitivity to the quark
distributions.  The size of $d(x)$ vs. $u(x)$ at high $x$ is one
of the remaining open questions for unpolarized quark
distributions, and if it can be established that the high
$x$---high $k$ results are mainly direct (or mainly
fragmentation, if that should happen against our expectations)
then the observed $\pi^+/\pi^-$ ratio becomes a direct measure
of $d(x)/u(x)$.

%%%%%%%%%%%%%%%%%%%
%%%%%%%%%%%%%%%%%%%

To elaborate the preceeding remarks, consider what happens for
$x$ approaching unity, where only valence quarks matter and
$\pi^+$ comes from $u$ and $\pi^-$ comes from $d$.  If
fragmentation dominated, then
\begin{equation}
\left. {\pi^+ \over \pi^-} \right|_{frag} 
                           = 4 \ {
           f_p u(x) + (1-f_p) d(x) \over
           f_p d(x) + (1-f_p) u(x)
                                 },
\end{equation}
where the target is fraction $f_p$ proton.   The ``4,'' of
course, is $(e_u/e_d)^2$.  Also, for fragmentation, $u(x)$ and
$d(x)$ must be understood as appearing inside some integrals, but
only the ratio $d(x)/u(x)$ as $x \rightarrow 1$ will matter here.

For the direct case, the short distance nature of the reaction
allows the photon to interact with the produced $q \bar q$ pair
as well as with the target quark, so it makes less difference
whether a $\pi^+$ or $\pi^-$ is produced.  One has---see
eqn.~(\ref{directeqn})---,
\begin{equation}
\left. {\pi^+ \over \pi^-} \right|_{direct} 
                           = 
      \left( s+2|u| \over 2s+|u| \right)^2 
                               {
           f_p u(x) + (1-f_p) d(x) \over
           f_p d(x) + (1-f_p) u(x)
                                 }.
\end{equation}
The prefactor is less than one, but approaches one for small
angles and maximum pion energy, when $|u| \rightarrow s$.

For isoscalar ($I_3 = 0$ suffices) or nearly isoscalar targets
the $\pi^+/\pi^-$ ratio would approach 4 for maximum momentum in
the fragmentation process, or approach 1 for the direct process,
and be rising with momentum.  One can qualitatively understand
the curve shown for the near isoscalar case in
Fig.~\ref{pionrat50}: at low $k$, fragmentation dominates but
the 
$\pi^+ / \pi^-$ ratio is not large as there are important
contributions from gluons and sea quarks in the target.  As $k$
rises, the valence quark contributions become relatively more
important and the ratio rises.  Then the direct or short
distance process takes over and the ratio falls, and finally
rises a bit due to the prefactor in the last equation after the
process is almost pure direct.

For a proton target, the $x \rightarrow 1$ limit of the ratio
$d(x)/u(x)$ is important.  Possibilities include,
\begin{eqnarray}
{d(x) \over u(x)} = \left\{
    \begin{array} {cl}
0   & \qquad {\rm many\ fits} \\
1/5 & \qquad {\rm pQCD} \\
1/2 & \qquad SU(6)
\end{array}
                              \right.
\end{eqnarray}
Both CTEQ~\cite{cteq95} and GRV~\cite{grv} have $u(x)$ and $d(x)$
falling with different powers of $(1-x)$, with $d(x)$ falling
faster, and so are examples of the first category.  The
BBS~\cite{bbs} distributions, whose non-separation of $q$ and
$\bar q$ is inconsequential at high
$q$, do satisfy the pQCD constraint and so give a different
$\pi^+ / \pi^-$ ratio as the pion momentum reaches its maximum.

%----------------------------------------------------------------

\section{Results with Polarization}  \label{polarized}

%----------------------------------------------------------------

We have calculated the asymmetry $E$ (or $A_{LL}$) for
$\pi^\pm$ photoproduction off both the proton and neutron.
If $R$ and $L$ represent photon helicities and $\pm$
represent target helicities, then $E$ is defined by
\begin{equation}
{E} \equiv {\sigma_{R+} - \sigma_{R-}
   \over
   \sigma_{R+} + \sigma_{R-} }.
\label{asym}
\end{equation}
The notation ``$E$'' comes from early pion
photoproduction work (see for example~\cite{barker}).
 We will now see how well we can
achieve our goal of determining the polarized parton
distributions from these experiments that measure $E$.  We have
direct sensitivity to the polarized gluon distribution $\Delta
g(x,\mu^2)$ since at moderate and lower momenta a reasonable
fraction of the pions are produced by reactions off the gluons
within the target.  Other determinations of $\Delta g$ have
depended upon higher order effects such as the evolution of
the polarized quark distributions~\cite{bfr96}, which is
driven in part by
$\Delta g$.

%%%%%%%%%%%%%%%%%%%
\begin{figure}

\epsfxsize 3.4 in \epsfbox{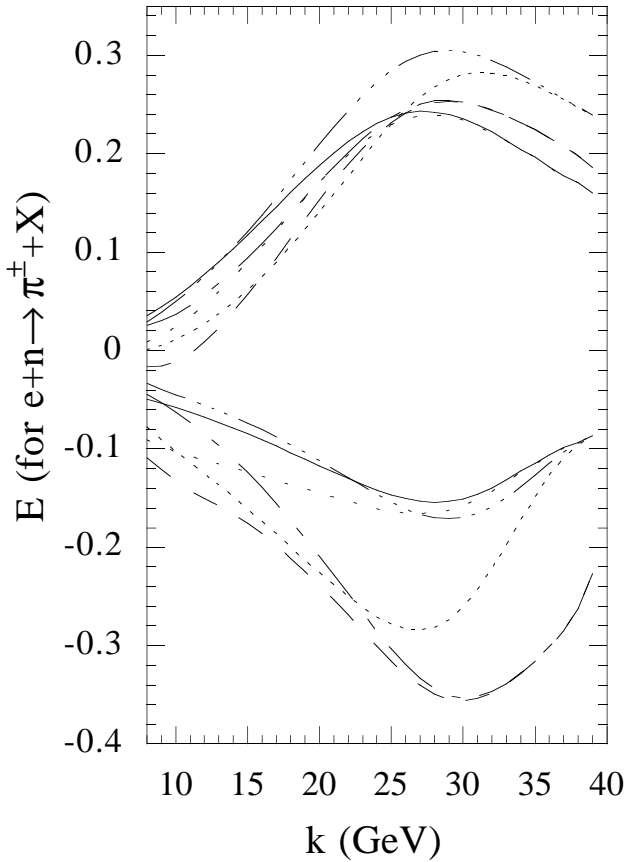}

\vglue 0.1 in

\caption{The asymmetry $E$ for 
$\vec \gamma + \vec n \rightarrow \pi^\pm +X$, at $E_e =
50$ GeV and $\theta_{lab} = 5.5^\circ$.   The upper six curves
are for $\pi^-$ production and the lower six curves are for
$\pi^+$ production.  For each set of six, there are three curves
with the full calculation, with the loose dotted curve using parton
distributions from GRSV, the dashed curve using GS-A, and the %%@
tight
dotted line using CTEQ/Soffer {\it et al.} and the BBS polarized
gluon distribution.  The other three curves have $\Delta g$ set to
zero, with the solid line  using GRSV, the
dash-dot curve using GS, and the dash-triple dot curve using
CTEQ/Soffer {\it et al}. }

\label{n-pi_50GeV}
\end{figure}
%%%%%%%%%%%%%%%%%%%

\begin{figure}

%\vglue  
\hskip 2.0 in \epsfxsize 3.4 in \epsfbox{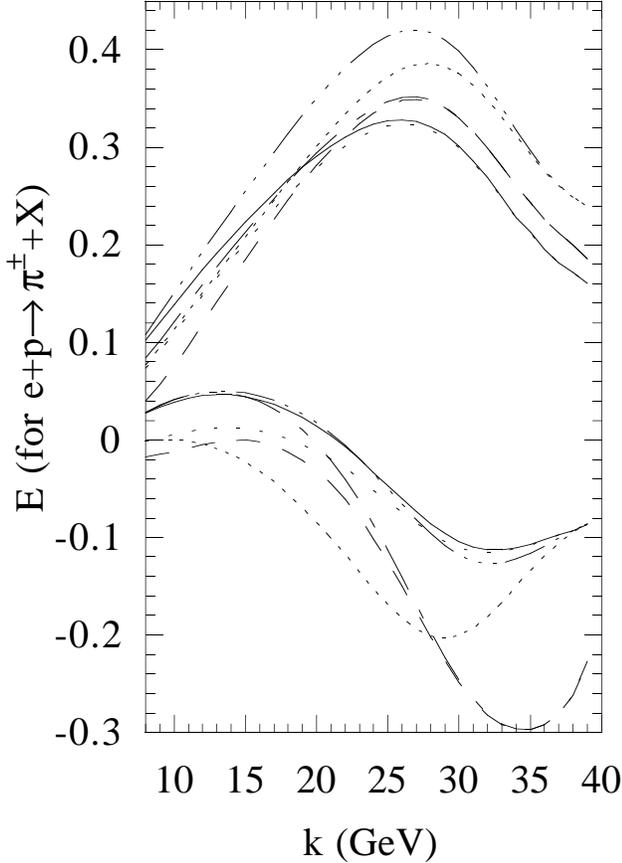}

\vglue 0.1 in

\caption{The asymmetry $E$ for 
$\vec \gamma + \vec p \rightarrow \pi^\pm +X$, at $E_e =
50$ GeV and $\theta_{lab} = 5.5^\circ$.  This time, the upper six
curves are for $\pi^+$ production and the lower six curves are for
$\pi^-$ production.  As in Fig.~\protect\ref{n-pi_50GeV}, for
each set of six, there are three curves with the full calculation,
with the loose dotted curve using parton distributions from GRSV, the
dashed curve using GS-A, and the tight dotted line using
CTEQ/Soffer {\it et al.} and the BBS polarized gluon
distribution.  The other three curves have
$\Delta g$ set to zero, with the solid         curve 
using GRSV, the dash-dot curve using GS, and the dash-triple dot
curve using CTEQ/Soffer {\it et al}.}

\label{p-pi_50GeV}
\end{figure}

We will begin by presenting results for 
$\vec \gamma + \vec n \rightarrow \pi^\pm +X$ and for 
$\vec \gamma + \vec p \rightarrow \pi^\pm +X$ where the
photon comes from radiation off an incoming electron beam
of energy $E_e = 50 GeV$ and the pions are observed at lab
angle $\theta_{lab} = 5.5^\circ$.  The outcome plotting
asymmetry $E$ vs. the magnitude of the pion momentum is
given in Figs.~\ref{n-pi_50GeV} and ~\ref{p-pi_50GeV} using
three differing sets of polarized parton models.  

The polarized parton models are those of Gehrmann and
Stirling (GS)~\cite{gs96}, of Gl\"uck, Reya, Stratmann, and
Vogelsang (GRSV)~\cite{grsv96}, and a suggestion of Soffer
{\it et al.}~\cite{soffer}.  Both the GS and GRSV polarized fits
use the fits of Gl\"uck, Reya, and Vogt (GRV)~\cite{grv} when
they need unpolarized distributions, at least in leading order.
%All fit the available data on $g_1$ from deep inelastic lepton
%scattering experiments.  
For the first two, we have obtained the renormalization scale
dependent results for the polarized parton distributions
directly from the authors. The Soffer {\it et al.}
suggestion relates the polarized and unpolarized distribution
functions, specifically,
\begin{eqnarray}
\Delta u_V(x) &=& u_V(x) - d_V(x) , \nonumber \\
\Delta d_V(x) &=& - {1\over 3} d_V(x) ,
\end{eqnarray}

\noindent and other polarized distributions are treated as
small.  When we use the Soffer {\it et al.}\ suggestion,
we team it with the CTEQ~\cite{cteq95} quark
distributions and the polarized gluon distribution of
Brodsky, Burkhart, and Schmidt (BBS)~\cite{bbs}.  In addition
this case requires a polarized distribution for the sea quarks,
which we take as
\begin{equation}
\Delta s(x) = -0.667 (1-x)^7,
\end{equation}
with the same sea distribution for up, down, and strange
quarks.  This gives 
$\langle \Delta s + \Delta \bar s \rangle = - 1/6$. In
all cases, we set the renormalization scale
$\mu^2$ to $k_T^2$, where $k_T$ is the transverse momentum of
the produced meson.

Although there are 12 different curves on
Fig.~\ref{n-pi_50GeV}, it is not so complicated. 
  In all cases,
$E$ is generally positive for the $\pi^-$ and negative for the
$\pi^+$, so there are six curves above for the $\pi^-$ and six
below for the
$\pi^+$.  Each of the three parton distribution models is
represented twice, once with the full calculation and once with
the polarized gluon distributions $\Delta g(x)$ (but not the total
gluon distribution $g(x)$) set to zero.  

Fig.~\ref{p-pi_50GeV}, for the proton target, is similar except
that $\pi^+$ is above and $\pi^-$ below.

One reaches the following conclusions from the graphs:

\begin{itemize}

\item At large pion momentum $k$ contributions from gluons
in the target are not significant but the results for differing
polarized quark distributions are quite different,  allowing the
data to discriminate among the various polarized quark
distribution models.  Note that for both the $\pi^\pm$ at high
$k$, two of the models give quite similar results and one is
different.  However, for the $\pi^-$ it is CTEQ/Soffer {\it et
al.} that is different, whereas for the $\pi^+$ it is GS that
stands out.

\item At low or moderate $k$ the results for the different
model polarized quark distributions are---if evaluated with
zero or the same $\Delta g$---rather similar. In the
figures, we show the curves with $\Delta g = 0$. The clearest
case is $\pi^-$ production off a proton target.   

\item At low or moderate $k$,  the differences among the
models are mainly due to the differences in $\Delta g$ (even
noting that the largest $\Delta g$ are not represented on these
two figures), and thus the measurements can discriminate among
the differing models for $\Delta g$.

\end{itemize}

%%%%%%%%%%%%%%%%%%%
%%%%%%%%%%%%%%%%%%%

\begin{figure}

%\vglue  
\hskip 2.0 in \epsfxsize 3.4 in \epsfbox{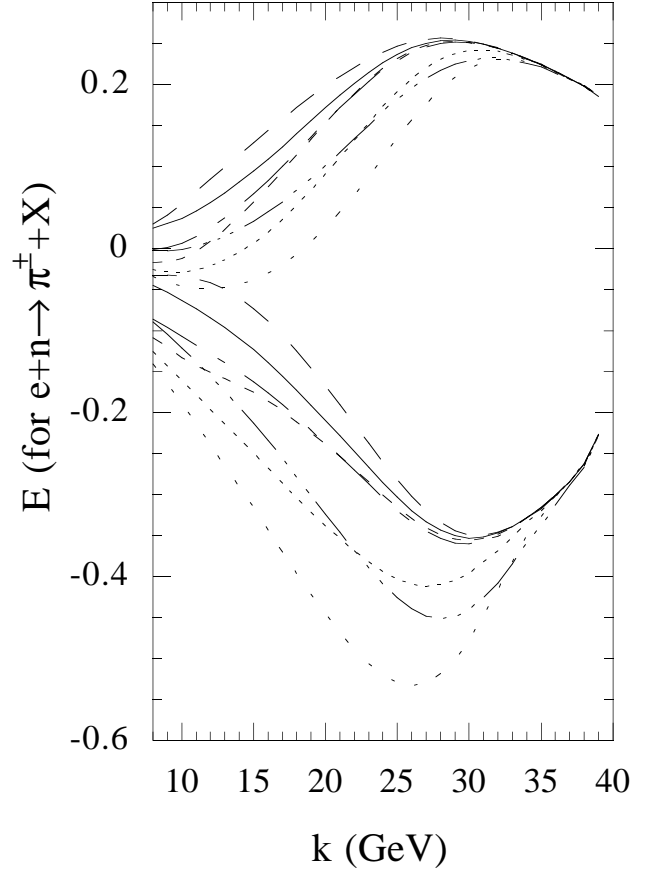}

\vglue 0.1 in

\caption{The asymmetry $E$ for 
$\vec \gamma + \vec n \rightarrow \pi^\pm +X$, at $E_e =
50$ GeV and $\theta_{lab} = 5.5^\circ$, with one model for the
quark distributions and several models for the polarized gluon
distribution.  We choose to use the quark distributions of GS. 
The solid curve is the benchmark with $\Delta g$ set to 
zero.  The short dashed curve uses the quark and unpolarized 
gluon distribution of GS but the polarized gluon distribution 
of GS model A.  The long dashed curve uses GS model C.  The dash
dot curve uses GRSV.  The dash triple dot curve similarly uses 
BBS, the tight dotted curve uses BFR model AR, and the loose 
dotted curve uses BFR model OS. }

\label{n-pi_50GeV_g}
\end{figure}

We elaborate on the last point in Fig.~\ref{n-pi_50GeV_g},
where we use only one quark distribution, but six different
gluon distributions to show the differences in their effect
upon this asymmetry.  Two of the new polarized gluon
distributions are from Ball, Forte, and Ridolfi
(BFR)~\cite{bfr96}, and we use versions AR and OS.  (Neither
BBS nor BFR give quark distributions for each individual flavor
quark and antiquark, so we can show results from their gluon
distributions only in combination with other authors's
models for the quarks.)  The other new polarized gluon 
distribution is GS version C.  One can see that the available
polarized gluon distributions, all inferred from $g_1$ data,
sometimes abetted by pQCD considerations at high
x~\cite{bbs},  give distinct results in the present case.

Incidentally, the minimum $x$ that enters the
calculation of the fragmentation process is the same as the
unique $x$ that enters the direct process, Eqn.~(\ref{x}). 
Hence for the situation of Figs.~\ref{n-pi_50GeV}
or~\ref{n-pi_50GeV_g}, the minimum $x$ for pion momentum
$k=8$ GeV is $x_{min} = 0.05$ and for $k=20$ GeV, 
$x_{min} = 0.16$.  This gives some idea of the $x$ range that
is probed by these experiments.

%%%%%%%%%%%%%%%%%%%
%%%%%%%%%%%%%%%%%%%

We continue showing results in Fig.~\ref{p-pi_50GeV} by
giving the analog of Fig.~\ref{n-pi_50GeV} but for  a proton
target.  The electron energy is still $E_e = 50$ GeV and
$\theta_{lab} = 5.5^\circ$.   The $\pi^-$ curves, which are the
lower ones in this Figure, bunch very well at low $k$ for the
three curves with $\Delta g$ set to zero, and the curves using the
$\Delta g$ pertinent to each model are quite distinct.  The
$\pi^+$ curves are less distinct from each other, but it is still
true that for the models chosen the curves with $\Delta g = 0$
all lie, at low $k$, above the curves with gluon polarization
included.

%%%%%%%%%%%%%%%%%%%
\begin{figure}

\epsfxsize 3.4 in \epsfbox{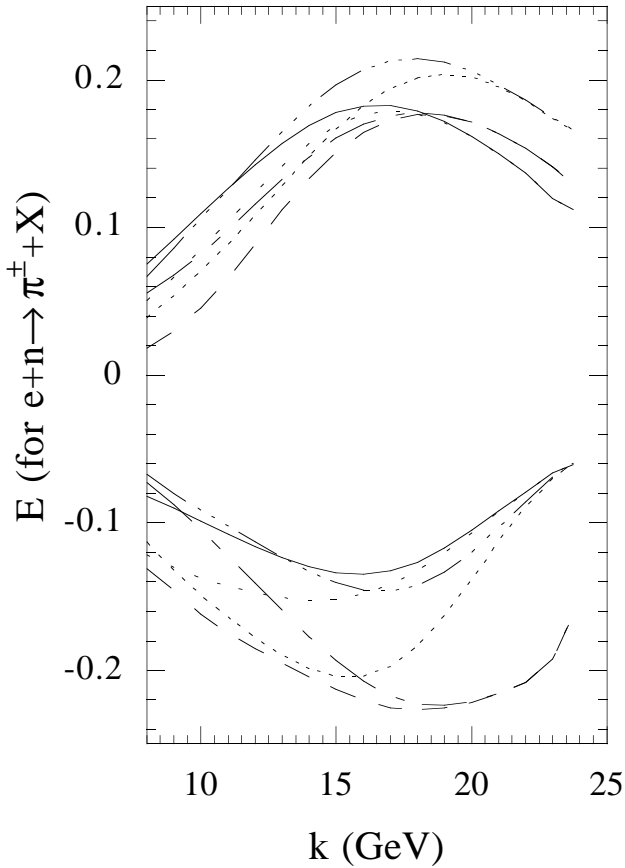}

\vglue 0.1 in

\caption{The asymmetry $E$ for 
$\vec \gamma + \vec n \rightarrow \pi^\pm +X$, at $E_e =
27.5$ GeV and $\theta_{lab} = 5.5^\circ$.  The remainder of the
caption is the same as for Fig.~\protect\ref{n-pi_50GeV}.}

\label{n-pi_27GeV}
\end{figure}
%%%%%%%%%%%%%%%%%%%

\begin{figure}

\epsfxsize 3.4 in \epsfbox{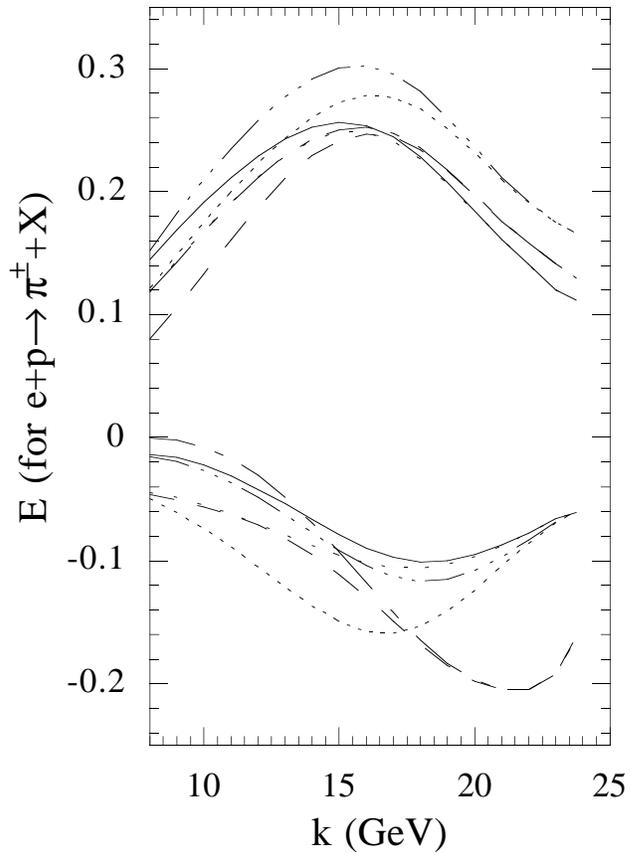}

\vglue 0.1 in

\caption{The asymmetry $E$ for 
$\vec \gamma + \vec p \rightarrow \pi^\pm +X$, at $E_e =
27.5$ GeV and $\theta_{lab} = 5.5^\circ$.  The remainder of the
caption is the same as for Fig.~\protect\ref{p-pi_50GeV}.}

\label{p-pi_27GeV}
\end{figure}

\begin{figure}

\epsfxsize 3.4 in \epsfbox{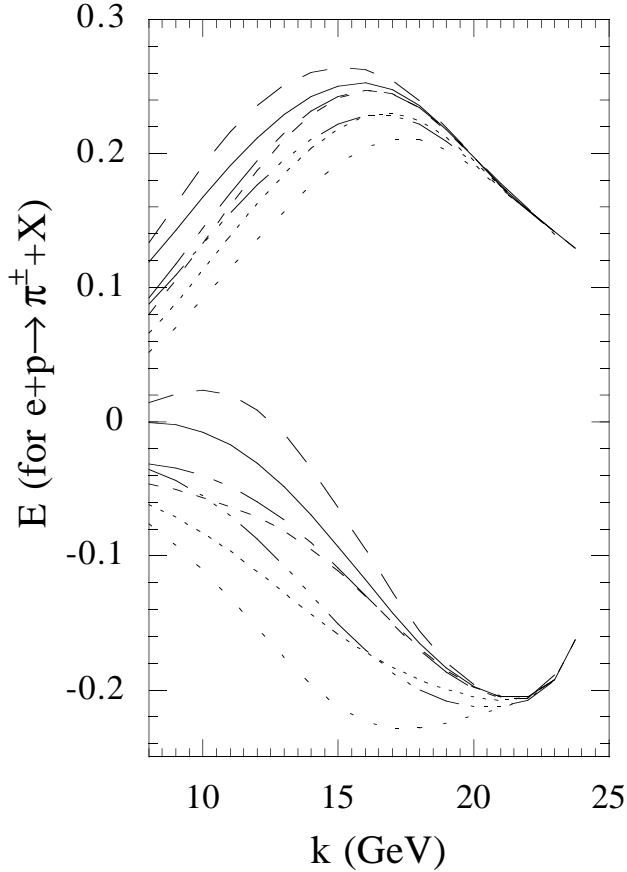}

\vglue 0.1 in

\caption{The asymmetry $E$ for 
$\vec \gamma + \vec p \rightarrow \pi^\pm +X$, at $E_e =
27$ GeV and $\theta_{lab} = 5.5^\circ$, with one model for the
quark distributions and several models for the polarized gluon
distribution.  We use the quark distributions of GS. 
The upper set of curves is for the $\pi^+$ and the lower set is %%@
for
the $\pi^-$;  otherwise the caption is the same as
Fig~\protect\ref{n-pi_50GeV_g}. }

\label{p-pi_27GeV_g}
\end{figure}

The next three figures show the analogs of the preceding three
Figures but for an incoming electron energy of 27.5 GeV; the lab
angle is still $\theta_{lab} = 5.5^\circ$.  Fig.~\ref{n-pi_27GeV}
shows the asymmetry $E$ for $\pi^\pm$ production off a neutron
target for the three models we have chosen, with and without
$\Delta g$.  Fig.~\ref{p-pi_27GeV} does the same for a proton
target.  The Figure with one quark distribution model but six
polarized gluon distribution models is Fig~\ref{p-pi_27GeV_g}. 
It is the analog of Fig.~\ref{n-pi_50GeV_g} , but for variation
we have given this Figure with a proton instead of a neutron
target.

\section{Discussion}      \label{discussion}

We feel we have demonstrated that with polarized initial states,
pion photoproduction at low (but still with $k_T$ above about 1
GeV) and moderate pion momenta can be a useful and successful way
to learn about the polarized gluon distribution.  In this region,
the various models for the polarized quark distributions all give
rather similar results when the effects of the polarized gluon
distributions are removed.  The effects of the polarized gluon
distributions are distinct for the different models, and
particularly for the BBS~\cite{bbs} and BFR~\cite{bfr96} models
are quite large.  For the kinematics we have looked at, the
resolved photon contributions are always small.  In the low to
moderate $k$ region, the fragmentation contribution is dominant.

At the highest allowed pion momentum the asymmetry does become
sensitive to the differences among the various quark models, and
so can empirically distinguish among them.  What we call the
direct process, i.e., pion production at short distance rather
than via fragmentation, dominates in this region.  In particular,
the high $x$ quarks of the target give the dominant contributions
and the models for the polarized quark distributions do not agree
at high $x$.

%%%%%%%%%%%%%%%%%%%
%%%%%%%%%%%%%%%%%%%

Questions may be asked about the use of perturbative QCD, upon
which our analyses depend.  We are, of course, mainly considering
ratios of cross sections so that many potential problems cancel
out.  

Also, studies of the polarized gluon distribution depend mainly
upon the fragmentation process.  This is a leading twist process,
so using perturbation theory to calculate it should be accurate
and has not generally been questioned.  

Within the context of perturbation theory, one may ask how
large the higher order in $\alpha_s$ corrections are.  For the
unpolarized case, the answer is that the next to leading
corrections double the result~\cite{aurenche}.  We should
state that we have simply doubled our lowest order calculations
to obtain our results: remember we are taking ratios.   
We are not aware of NLO calculations of 
$\vec \gamma \vec g \rightarrow q \bar q$ or 
$\vec \gamma \vec q \rightarrow q g$.  
However, NLO calculations of 
$\vec g \vec q \rightarrow \gamma q $ and
$\vec q {\vec {\bar q}} \rightarrow \gamma g$ and related 
$2 \rightarrow 3$ processes have been done~\cite{andy}.
The $K$-factors [ratio of LO + NLO to LO cross sections] for the
polarized cross section $\Delta \sigma$ always exceed unity, so
that the effect of the NLO corrections upon the ratio $E$ for
direct photon production is not great.

%%%%%%%%%%%%%%%%%%%
%%%%%%%%%%%%%%%%%%%

Much of our further discussion concerns the direct process, which 
is a higher twist contribution and using perturbative QCD has been
questioned is some such cases.  We can for completeness summarize
some earlier arguments~\cite{acw1}.

Our analysis requires that the ``$X$'' in 
$\gamma + p \rightarrow \pi + X$ is  out of the resonance
region.  For the energies we have considered, this is easy to
satisfy except at the highest $k$.

Another question regards further higher twist corrections, for
example, corrections due to the quarks in the pion
having finite momentum transverse to the pion's
overall momentum.  This has been much studied in the
context of the pion electromagnetic form
factor~\cite{listerman}.  In the present case, the virtual gluon
in the direct process is much farther off shell~\cite{acw1,cw93}
than for the pion form factor at presently accessible kinematics.
Hence higher twist effects will be less significant for measurable
photoproduction of high transverse momentum mesons than for meson
form factors at any currently measured momentum transfers.
 Similarly, we have not considered transverse momentum
smearing of the incoming quarks.  It has been considered in the
context of pion production in $pp$ and $\bar pp$ collsions, and
does have some effect there on the extraction of the polarized
gluon distribution~\cite{transversepp}.

Perturbative corrections that are higher order in
$\alpha_s$ have not been calculated.  They may be
calculated along the lines of~\cite{braaten83} for the 
$\pi^0\gamma\gamma$ and of~\cite{braatentse} for the 
$\pi^\pm$ electromagnetic form factors.  For both of
these, using the asymptotic distribution amplitude and a
suitable choice of renormalization scale, the magnitude
of the correction was about 20\%, decreasing the
$\pi^0\gamma\gamma$ and increasing the $\pi^\pm$ form
factors.

Our calculations can also be applied to production of kaons and
to neutral pions.  For neutral pions, the fragmentation process
cross section is the average of the $\pi^+$ and $\pi^-$
cross sections.  However, the direct production of neutral pions
is less than the direct production of either charged pion.  Useful
studies are also possible using single polarization asymmetries.  
We hope to return to these subjects in the near future.

We conclude with a summary.

\begin{itemize}

\item Of the three processes that contribute to high $k_T$ pion
production, or to its single arm electron equivalent 
$e + N \rightarrow \pi + X$, the resolved photon process is
unimportant for incoming energies of a few 10's of GeV and small
angles.  The fragmentation process dominates at low or moderate
momenta, the direct process dominates at high pion momenta.

\item The $\pi^+/\pi^-$ ratio predictions are different for
fragmentation and direct processes.  For isoscalar or near
isoscalar targets, with pions at high momenta, fragmentation
would give about a 4:1 ratio [coming from $(e_u/e_d)^2$] but the
direct process gives about 1:1.  Verification that short
distance production takes over from fragmentation production
lies in seeing a fall in the $\pi^+/\pi^-$ ratio (still for
$I=0$ targets) as the takeover occurs.

\item Where the direct process dominates, and without
polarization, the rate is proportional to things that are known
and the $I_\pi$, the same integral over the pion distribution
amplitude that fixes $\gamma^* + \gamma \rightarrow \pi^0$ and
$\gamma^* + \pi^\pm \rightarrow \pi^\pm$.  Currently data for
the last two processes taken at face value gives discordant
values of $I_\pi$.

\item  With initial state polarization, one can form the double
helicity asymmetry $A_{LL}$ or $E$.  Where the direct process
dominates, at high pion momentum, the asymmetry is proportional
to the polarized quark distributions $\Delta u$ for the $\pi^+$
and $\Delta d$ for the $\pi^-$, times things that are known or
easily calculable.  Hence, one can measure the $\Delta q_i$
individually.

\item When the fragmentation process dominates, experiments with
initial state polarization are sensitive to $\Delta g$.  The
polarization asymmetry is 100\% in magnitude for the production
off a gluon target, and the current spectrum of models for
$\Delta g$ leads to a wide diversity of $A_{LL}$ or $E$
predictions for pion photoproduction in the fragmentation region.

\end{itemize}

\section*{acknowledgments}

We thank V. Breton, J. Gomez, and K. Griffioen for useful
discussions and T. Gehrmann, M. Stratmann, and J. Qiu for
supplying parton distribution computer codes from the GS,
GRSV, and CTEQ collaborations, respectively. AA thanks the
DOE for support under grant DE--AC05--84ER40150; CEC and CW
thank the NSF for support under grant PHY-9600415.

%------------------------------------------------
\end{document}